# Two Scenarios of Breaking Chaotic Phase Synchronization

A. A. Koronovskii, M. K. Kurovskaya, O. I. Moskalenko, and A. E. Hramov

*Saratov State University, Saratov, 410012 Russia*

**Abstract**—Two types of phase synchronization (accordingly, two scenarios of breaking phase synchronization) between coupled stochastic oscillators are shown to exist depending on the discrepancy between the control parameters of interacting oscillators, as in the case of classical synchronization of periodic oscillators. If interacting stochastic oscillators are weakly detuned, the phase coherency of the attractors persists when phase synchronization breaks. Conversely, if the control parameters differ considerably, the chaotic attractor becomes phase-incoherent under the conditions of phase synchronization break.

## INTRODUCTION

The phenomenon of phase synchronization observed in systems of various nature [1, 2], including chemical, biological, and physiological systems, is being today attracting much interest of researchers [3–5]. One point needs to be made at once. Initially Russian authors meant the term "phase synchronization" (which originated rather long ago) as the synchronization of oscillators the phases of which are converted in any event and then "applied" to them again [6–8], the oscillators and phase converters may be quite unlike. The principle of phase synchronization thus understood underlies the operation of electronic self-tuning systems, synchronous machines, phase electric drives, etc.

Today, the term "phase synchronization" (see, e.g., [3, 9–12]) is used in terms of works [13–15], which, of course, leads to confusion. Note, therefore, that, in this work, we see the term "phase synchronization" in terms of [3, 13–16].

The concept of phase synchronization is based on the notion of instantaneous phase $\varphi(t)$ of a chaotic signal [13, 14, 16]. Note at once that a universal approach of defining the chaotic signal phase that provides correct results for any dynamic system is lacking. Some ways of phase definition are appropriate to "good" systems with a simple topology of the strange attractor, which are called "systems with a well-defined phase" or "systems with a phase-coherent attractor." The chaotic attractor of such systems must be such that the projection of the phase trajectory onto some plane of states $(x, y)$ constantly rotates about the origin without intersecting and bending round it. In this case, instantaneous phase $\varphi(t)$ of a chaotic signal may be introduced as an angle on plane $(x, y)$ in the polar coordinate system [15, 17],

$$\varphi = \arctan\frac{y}{x}. \quad (1)$$

The fact that the projection of the phase trajectory all the time rotates around the origin without bending round it leads to the situation when mean frequency $\Omega$ of a chaotic signal, which is defined as the mean rate of change of the phase,

$$\Omega = \lim_{t \to \infty}\frac{\varphi(t)}{t} = \langle\dot\varphi(t)\rangle, \quad (2)$$

coincides with fundamental frequency $\Omega_0 = 2\pi f_0$ of Fourier spectrum $S(f)$ of system oscillations. Such a situation is considered as a proof that the chaotic signal instantaneous phase thus introduced is correct [18]. If, however, the phase trajectory projection bends round the origin from time to time (i.e., does not make a full turn around), the origin is shaded by parts of the phase trajectory. Such an attractor is called "phase-incoherent" and the system itself, "system with an ill-defined phase."

Another way of defining the phase of a chaotic dynamic system consists in introducing into consideration the analytical signal [13, 16]

$$\zeta(t) = x(t) + jH[x] = A(t)e^{j\varphi(t)}, \quad (3)$$

where function $H[x]$ is the Gilbert transform of time realization $x(t)$,

$$H[x] = \frac{1}{\pi}\text{v.p.}\int_{-\infty}^{+\infty}\frac{x(\tau)}{t-\tau}d\tau \quad (4)$$

(*vp* means that the integral is taken in the sense of the principal value). Accordingly, phase φ(*t*) of chaotic signal *x*(*t*) is determined from relationships (3) and (4).

The third standard way of defining the chaotic signal phase is introduction of the Poincaré section surface [13, 16], in which case he phase is defined as

$$\varphi(t) = 2\pi \frac{t - t_n}{t_{n+1} - t_n} + 2\pi n, \quad t_n \leq t \leq t_{n+1}, \quad (5)$$

where $t_n$ is the time instant the phase trajectory intersects the Poincaré surface.

Clearly, ways (1) and (5) yield nearly the same results for the phase-coherence attractor (in these cases, the dynamics of the instantaneous phase will differ slightly over time intervals shorter than the characteristic time taken for the trajectory to return to the Poincaré surface) [3]. It is known that the instantaneous phase introduced by the Gilbert transform behaves for the phase-coherent attractor virtually in the same manner as the phases defined by (1) and (5) (see, e.g., [15]).

The phase synchronization condition implies that phases $\varphi_{1,2}(t)$ of chaotic signals from interacting oscillators (or the phases of a stochastic oscillator and an external signal) become locked: their difference does not grow in absolute value with time (i.e., does not exceed some constant assigned in advance),

$$|\Delta\varphi(t)| = |\varphi_1(t) - \varphi_2(t)| < \text{const.} \quad (6)$$

It is obvious that phase lock-in causes coincidence of the frequencies of the chaotic signals, which, as follows from relationships (2) and (6), must be the same for the interacting systems.

As a rule, no distinction is made between the phase synchronization regimes. The only exception is work [19], where an attempt was made to substantiate the existence of three different ways of transition to phase synchronization according to the system properties. In this work, we show that there are only two types of phase synchronization between stochastic oscillators, which depend on the mismatch between the control parameters of interacting systems (exactly as in the case of classical synchronization of periodic oscillators). Accordingly, there are two scenarios of breaking the synchronous dynamics of interacting systems. If the mismatch between the parameters of interacting oscillators is small, their chaotic attractors arising from synchronization breaking will remain phase-coherent. If the mismatch is large, then, as the parameter of coupling between interacting systems (or the amplitude of an external action) decreases, at least one of the attractors of interacting oscillators loses phase coherency (i.e., becomes phase-incoherent) below the threshold of synchronization breaking (onset).

## 1. SYNCHRONIZATION OF PERIODIC OSCILLATIONS IN TERMS OF THE CONCEPT OF CHAOTIC PHASE SYNCHRONIZATION

Consider the behavior of a nonautonomous van der Pol oscillator (a reference model in the nonlinear theory of oscillations) subjected to a harmonic action in terms of the theory of chaotic phase synchronization for the cases of small and large frequency mismatches between the external forcing signal and eigenmode. This system is described by the equation

$$\ddot{x} - (\lambda - x^2)\dot{x} + x = A\sin(2\pi f_e t), \quad (7)$$

where $A$ and $f_e$ are, respectively, the amplitude and frequency of the harmonic external action and $\lambda$ is the nonlinearity parameter. The given system is the well-known classical model of the synchronization theory, and its behavior is described even in manuals of physics (see, e.g., [20, 21]). In the case at hand, it is clear that oscillations are not chaotic; however, the notions of signal phase, chaotic phase synchronization, and attractor phase coherence are readily applicable in this case too, as is done in considering the behavior of stochastic oscillators.

The nonautonomous behavior ($A \neq 0$) of system (7) strongly depends on the mismatch between the oscillation eigenfrequency and the frequency of the external action, as well as on the amplitude of the external action. It is well known [16, 20, 21] that, when the frequency mismatch is small or large, the synchronization condition sets in variously with increasing mismatch. In the former case, the frequency of the self-oscillator is locked by the external force; in the latter, the self-oscillation frequency is suppressed and the intensity of oscillations at the frequency of the forcing force grows. When nonlinearity parameter $\lambda$ is small, one can analyze the dynamics of system (7) with the method of slowly varying amplitudes by passing to truncated equations for the oscillation complex amplitude. While for a small mismatch synchronization shows up as a saddle-node bifurcation on the complex amplitude plane, for large mismatches synchronization break appears as a series of Andronov–Hopf bifurcations in the vicinity of a stationary stable point and the limit cycle amplitude grows until the cycle crosses the origin (for more details, see [16, 21]). At a large frequency mismatch, the synchronization regime, as applied to phase synchronization, is violated exactly when the limit cycle crosses the origin (at a small mismatch, the phase synchronization condition and frequency lock condition are violated simultaneously). At the instant the limit cycle crosses the origin on the complex amplitude plane, the corresponding attractor is no longer phase-coherent. Therefore, when periodic self-oscillations lock in synchronism due to an external harmonic action at a large frequency mismatch, the violation of phase synchronization is equivalent to the violation of attractor phase coherence.

Thus, when a frequency mismatch is large, the attractor of the van der Pol oscillator becomes phase-incoherent in the domain lying below the phase synchronization tongue on the parameter plane $(f_e, A)$, while at a small mismatch, the attractor is phase-coherent both inside and outside the synchronization tongue. The corresponding situation is depicted in Fig. 1, which shows the phase portrait on the plane of states $(x, \dot{x})$ and plots the time variation of phase difference $\Delta\varphi$ between the external signal and the time realization of the nonautonomous system above and below the phase synchronization threshold for small (Figs. 1a, 1b) and large (Figs. 1c, 1d) frequency mismatches. When the mismatch is small (Figs. 1a, 1b), the attractor retains phase coherence when phase synchronization breaks (when phase lock condition (6) becomes invalid). When the mismatch is large (Figs. 1c, 1d), condition (6) becomes invalid, since the attractor ceases to be phase-coherent. It can be shown that the same pattern will also be observed for the stochastic oscillator subjected to an external harmonic action. We, however, will consider more complicated cases yet demonstrating the same two scenarios of phase synchronization break.

## 2. BREAK OF CHAOTIC PHASE SYNCHRONIZATION IN A SYSTEM OF TWO UNIDIRECTIONALLY COUPLED RESSLER OSCILLATORS

Consider now a more complicated example of phase synchronization break in a system of two unidirectionally coupled stochastic Ressler oscillators, which feature phase-coherent chaotic attractors in the autonomous regime. The dynamics of such a system is described by a set of differential equations

$$\dot{x}_d = -\omega_d y_d - z_d, \quad \dot{x}_{dr} = -\omega_{dr} y_{dr} - z_{dr} + \varepsilon(x_d - x_{dr}),$$
$$\dot{y}_d = \omega_d x_d + a y_d, \quad \dot{y}_{dr} = \omega_{dr} x_{dr} + a y_{dr}, \quad (8)$$
$$\dot{z}_d = p + z_d(x_d - c), \quad \dot{z}_n = p + z_{dr}(x_{dr} - c),$$

where $\varepsilon$ is the coupling parameter. The values of the control parameters are the following [22]: $a = 0.15$, $p = 0$, and $c = 10.0$. Parameter $\omega_{dr}$ specifying the eigenfrequency of the driven subsystem was taken to be $\omega_{dr} = 0.95$; the same parameter for the driving subsystem, $\omega_d$, was varied from 0.8 and 1.1 so as to mismatch (detune) the interacting oscillators. As was noted above, with control parameters $a$, $p$, $c$, and $\omega_{dr}$ thus chosen, the chaotic attractors of both subsystems when uncoupled are phase-coherent throughout the range of $\omega_{dr}$.

Figure 2 shows the parameter plane $(\omega_d, \varepsilon)$ for system (8). The continuous line marks the boundary of the phase synchronization regime. The presence or absence of synchronization depends on whether the phase lock condition is fulfilled (for details, see [3–5, 16]). The instantaneous phase of the chaotic signal was routinely defined as the rotation angle on the plane of variables $(x, y)$: $\varphi = \arctan(y/x)$. Such an approach is valid, since the chaotic attractors are phase-coherent.

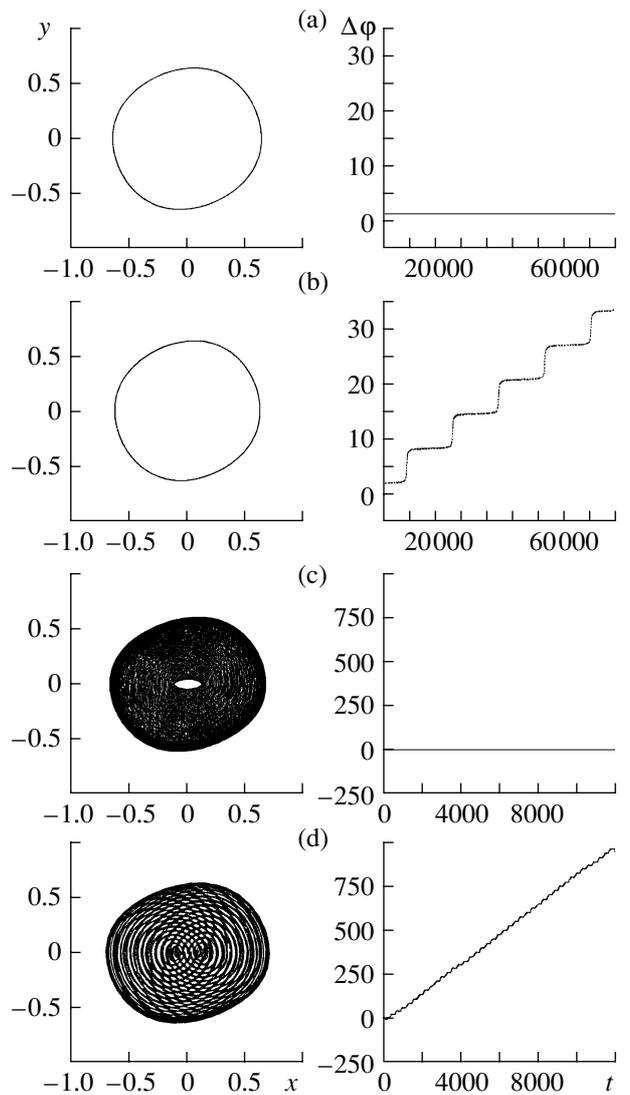

**Fig. 1.** Attractors of the nonautonomous van der Pol generator subjected to a harmonic signal and the time dependence of phase difference $\Delta\varphi(t) = \varphi(t) - 2\pi f_e t$ at (small) frequency mismatch $f_e = 0.1583$ (a) above $(A = 0.0065)$ and (b) below $(A = 0.0055)$ the phase synchronization boundary on the parameter plane $(f_e, AA)$. Phase $\varphi(t)$ of the signal is defined as rotation angle $\varphi = \arctan(y/x)x$ on the plane $(x, y)$, where $y(t) = \dot{x}(t)$. Control parameter $\lambda$ equals 0.1, and the eigenfrequency of the van der Pol generator is $f_0 = 0.1592$. Shown also are the same pictures for a larger mismatch $(f_e = 0.1275)$ (c) above $(A = 0.1450)$ and (d) below $(A = 0.1400)$ the phase synchronization boundary.

When investigating into the behavior of two unidirectionally coupled Ressler systems, we discovered two scenarios of phase synchronization breakdown, as in the case when periodic oscillations are synchronized by applying an external harmonic force. Figures 3a and 3b illustrate the chaotic attractors of the driven subsystem and the dependence $\Delta\varphi(t) = \varphi_d(t) - \varphi_{dr}(t)$ above and below the phase synchronization boundary when parameters $\omega_d$ and $\omega_{dr}$ are close to each other (small mismatch; in Fig. 2, the phase synchronization boundary at a small mismatch is shown by the thick continuous line). It is easy to see that, as coupling parameter $\varepsilon$ decreases, the phase synchronization regime breaks although the chaotic attractors remain phase-coherent (Fig. 3b). Figures 3c and 3d demonstrate the second scenario of phase synchronization breakdown for the interacting subsystems, which takes place at a large mismatch between the control parameters. Below the phase synchronization boundary, the chaotic attractor of the Ressler driven subsystem becomes phase-coherent and the phase synchronization condition is therefore violated. Clearly, the breakdown mechanisms in both cases are the same as in the case of periodic oscillations (cf. Figs 1 and 3) and can be explained by interaction between basic components in the power spectra of coupled stochastic oscillators.

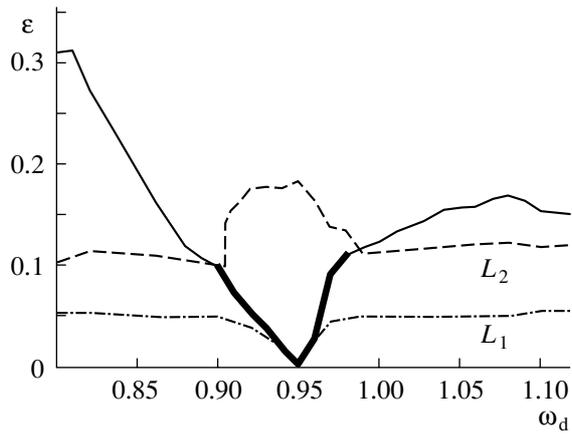

**Fig. 2.** Phase synchronization boundary on the parameter plane ($\omega_d$, $\varepsilon$) for system (8). In the range of small (large) mismatches $|\omega_{dr} - \omega_d|$ between the control parameters of coupled oscillators, the boundary is shown by the thick (thin) continuous line. Dashed line $L_1$ corresponds to the passage of one of the zero Lyapunov exponents into the negative domain, and dashed line $L_2$ corresponds to vanishing of one positive Lyapunov exponent.

## 3. HIGHEST LYAPUNOV EXPONENTS AND THE BOUNDARY OF PHASE SYNCHRONIZATION

The phase synchronization condition is often described with Lyapunov exponents (see, e.g., [3, 19]). It is therefore reasonable to see how variations of the Lyapunov exponents (which vary with the coupling parameter) are related to the occurrence of phase synchronization.

The system being considered (two unidirectionally coupled Ressler oscillators, see (8)) is characterized by six Lyapunov exponents, three of which, $\lambda_{d1} > \lambda_{d2} > \lambda_{d3}$, describe the dynamics of the driving subsystem and therefore are independent of coupling parameter $\varepsilon$]. The other three (conditional Lyapunov exponents [23, 24]), $\lambda_{dr1} > \lambda_{dr2} > \lambda_{dr3}$, describe the dynamics of the driven subsystem and grow with the degree of coupling between the subsystems (i.e., with $\varepsilon$). Clearly, at $\varepsilon = 0$, the systems are autonomous and so the Lyapunov exponent spectra are bound to have two zero exponents, $\lambda_{d2}$ and $\lambda_{dr2}$, being related to the driving and driven oscillators, respectively.

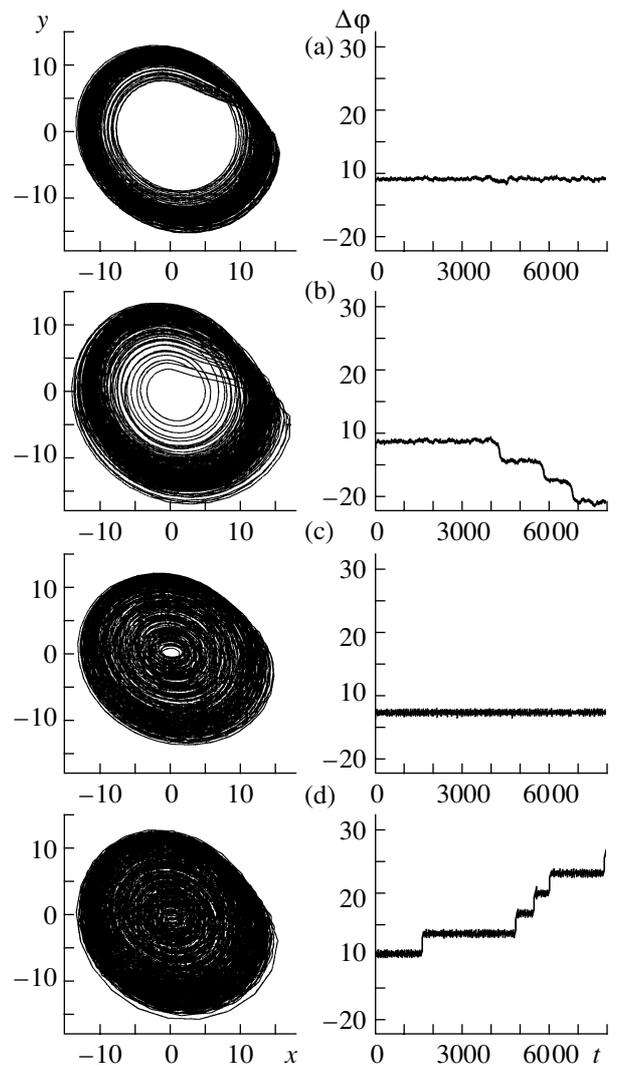

**Fig. 3.** Attractors of the driven subsystem in Ressler system (8) and the time dependence of phase difference $\Delta\varphi(t) = \varphi_d(t) - \varphi_{dr}(t)$ at a small frequency offset from the parameters of the driving subsystem ($\omega_d = 0.91$) (a) above ($\varepsilon = 0.077$) and (b) below ($\varepsilon = 0.075$) the phase synchronization boundary. Shown also are the same pictures for a larger mismatch ($\omega_d = 1.00$) (c) above ($\varepsilon = 0.125$) and (d) below ($\varepsilon = 0.123$) the phase synchronization boundary.

Figure 4 plots the four highest Lyapunov exponents for coupled Ressler oscillators (8) against the coupling parameter at $\omega_d = 0.93$. The exponent spectrum was cal-culated using the Benettin procedure with Gramm–Schmidt orthogonalization [25]. It is distinctly seen that Lyapunov exponents $\lambda_{d1,2}$ (dashed lines) do not vary with $\varepsilon$ unlike two others ($\lambda_{dr1,2}$, continuous lines). The lowest exponents, $\lambda_{d3}$ and $\lambda_{dr3}$, are on the order of $-10$ and so have a negligible effect on the processes in the system.

At a certain value of $\varepsilon$, one of the zero Lyapunov exponents, namely, $\lambda_{dr2}$, becomes negative. It is generally believed that the transition of the zero Lyapunov exponent to the negative range is a direct manifestation of the phase synchronization regime provided that the attractors of the interacting oscillators are phase-coherent in the absence of coupling (for details, see [3, 17, 19]). From Fig. 4 it follows that this Lyapunov exponent becomes negative somewhat earlier than phase synchronization sets in. In fact, the value of the coupling parameter corresponding to the onset of phase synchronization, $\varepsilon_{ps}$, is other than $\varepsilon$ at which one of the zero Lyapunov exponents ($\lambda_{dr2}$) pass into the negative range. As the coupling parameter grows, one of positive Lyapunov exponents ($\lambda_{dr1}$) also goes into the negative domain (Fig. 4).

Figure 2 shows, along with the phase synchronization boundary on the parameter plane ($\omega_d, \varepsilon$)), two characteristic curves $L_1$ and $L_2$, which are relevant the behavior of the Lyapunov exponents for system (8). The first curve, $L_1$, corresponds to $\varepsilon$ at which one of the zero Lyapunov exponents ($\lambda_{dr2}$) passes into the negative range. The second one, $L_2$, reflects the situation when one of the positive Lyapunov exponents ($\lambda_{dr1}$ in our case) passes through zero.

The results obtained in this work refute the assertion that the passage of a zero Lyapunov exponent into the negative domain is a direct indication of the phase synchronization regime in the case of interacting stochastic oscillators with phase-coherent attractors. As is easily seen from Fig. 2, curve $L_1$ coincides with the phase synchronization boundary only in a very narrow range of the mismatch between the parameters of coupled oscillators, although the attractors of the driving and driven subsystems in the autonomous regime are phase-coherent. Moreover, even if the mismatch is small, one cannot say that the synchronization boundary is indicated by the passage of a zero Lyapunov exponent into the negative domain. Indeed, if, for example, $\omega_d = 0.9$ (this value is a line of demarcation between the two types of phase synchronization breakdown), the coupling parameter at the synchronization boundary, $\varepsilon_{ps} \approx 0.099$, is twice as large as $\varepsilon_{L_1}$ corresponding to line $L_1$ ($\varepsilon_{L_1} \approx 0.05$) (see also Fig. 2). Thus, it follows from the above comparison that the passage of one zero Lyapunov exponent into the negative domain precedes the phase synchronization, rather than manifests the onset of the synchronous dynamics. It can be supposed that such a passage is more likely related to the synchronization of time scales [26–28]; however, this point calls for further investigation.

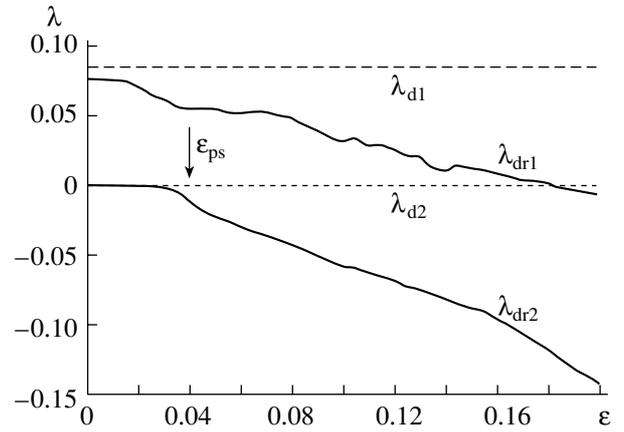

**Fig. 4.** Lyapunov exponents of two unidirectionally coupled Ressler oscillators (system (8)) vs. the coupling parameter. Dashed curves $\lambda_{d1}$ and $\lambda_{d2}$ correspond to the Lyapunov exponents for the driving system; continuous curves $\lambda_{dr1}$ and $\lambda_{dr2}$, to those for the driven system. The least exponents, $\lambda_{d3}$ and $\lambda_{dr3}$, are omitted, since they are on the order of $-10$ and have a negligible effect on the problem. The marked value of the coupling parameter, $\varepsilon_{ps}$, indicates the onset of phase synchronization.

In [19], curve $L_2$ on the plane ($\omega_d, \varepsilon$), which corresponds to vanishing of the positive Lyapunov exponent, was used to discriminate various scenarios of transition to phase synchronization depending on the coherence of the chaotic attractors of interacting systems. If the phase synchronization boundary runs above $L_1$(i.e., $\varepsilon_{ps} < \varepsilon_{L_2}$), the cases $\varepsilon_{ps} > \varepsilon_{L_2}$ (curve $L_2$ lies above the phase synchronization boundary) and $\varepsilon_{ps} > \varepsilon_{L_2}$ (curve $L_2$ lies below the phase synchronization boundary) are assumed to meet different types of transition to the phase synchronization regime, and this difference is due to the properties of chaotic attractors. According to [19], it is generally believed that the case $\varepsilon_{ps} < \varepsilon_{L_2}$ corresponds to the synchronization of oscillators with phase-incoherent chaotic attractors, while the case $\varepsilon_{ps} > \varepsilon_{L_2}$ is typical of coupled systems with heavily incoherent attractors. Nevertheless, Fig. 2 shows that both cases may also be observed in coupled systems with initially phase-coherent attractors. Moreover, the position of curve $L_2$ on the plane ($\omega_d, \varepsilon$) is in no way related to phase synchronization: it is completely dependent on the mechanisms described in [22, 29].

Based on the aforesaid, we can argue that the conclusion about three scenarios of transition to phase synchronization that has been reached [19] by analyzing the behavior of the highest Lyapunov exponents is invalid.

Thus, for two unidirectionally coupled Ressler oscillators defined by system (8), as well as for nonau-tonomous van der Pol generator (7), only two scenarios of transition to the phase synchronization regime should be distinguished. Which of them takes place depends on the control parameter mismatch.

## 4. INTERACTION OF BIDIRECTIONALLY COUPLED STOCHASTIC RESSLER OSCILLATORS

Let us show that the same two types of synchronization break are observed at bidirectional coupling of stochastic oscillators. To this end, consider two mutually coupled Ressler systems described in [19],

$$\dot{x}_{1,2} = -\omega_{1,2}y_{1,2} - z_{1,2},$$
$$\dot{y}_{1,2} = \omega_{1,2}x_{1,2} + ay_{1,2} + \varepsilon(y_{2,1} - y_{1,2}), \quad (9)$$
$$\dot{z}_{1,2} = 0.1 + z_{1,2}(x_{1,2} - 8.5),$$

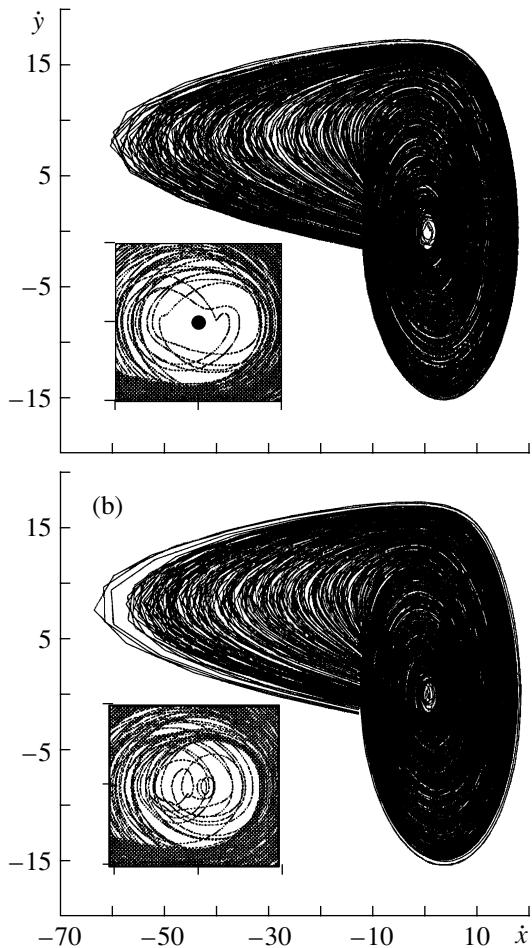

**Fig. 5.** Chaotic attractors of the first Ressler oscillator in system (9) on the plane $(\dot{x}, \dot{y})$ for $a = 0.21$. (a) Phase synchronization regime ($\varepsilon] = 0.055$), the attractor on the plane $(\dot{x}, \dot{y})$ is phase-coherent, and (b) phase-incoherent attractor when phase synchronization is not detected $\varepsilon] = 0.05$). The insets are the enlarged views of the trajectories on the plane $(\dot{x}, \dot{y})$ near the origin. In the inset to panel (a), the trajectory is seen to rotate about zero marked by the black dot.

with the same set of control parameters. In (9), $\varepsilon$ serves as a coupling parameter, $\omega_1 = 0.98$, and $\omega_2 = 1.02$.

System (9) is of interest in that one, analyzing its behavior, can tackle the question as to which types of phase synchronization will be observed for two oscillators with initially phase-incoherent chaotic attractors. The point is that, in the autonomous regime, both oscillators may have both phase-coherent and phase-incoherent attractors depending on parameter $a \in [0.15; 0.3]$, which defines the topology of the chaotic attractor. Such a situation was considered, in particular, in [19, 30], where interacting systems initially, when autonomous, were characterized by phase-incoherent chaotic attractors. In order to introduce the phase of the chaotic signal and then apply the methods from the phase synchronization theory, the authors of those works passed from the vector of states $(x, y, z)$ to the vector of velocities $(\dot{x}, \dot{y}, \dot{z})$. With such a transition, the attractor on the velocity plane $(\dot{x}, \dot{y})$ becomes phase-coherent, thereby making it possible to apply standard approaches of the phase synchronization theory.

When parameter $a$ exceeds some critical value $a_c$ ($a_{c1} \approx 0.186$ for $\omega_1$ and $a_{c2} \approx 0.195$ for $\omega_2$), the chaotic attractor of the autonomous Ressler system becomes phase-incoherent [19] and phase $\varphi(x)$ is defined as rotation angle $\varphi = \arctan(\dot{y}/\dot{x})$ on the velocity plane $(\dot{x}, \dot{y})$. Such a definition, as was already noted, allows one to diagnose the phase synchronization regime in systems with a spiral attractor. Having studied the behavior of two mutually coupled Ressler systems described by (9) along the phase synchronization boundary on the parameter plane $(a, \varepsilon)$ found in [19], we again discovered two scenarios of synchronization break, as in the case of unidirectionally coupled oscillators described by (8). At a small value of $a$ ($a < a_* = 0.205$), the first type of synchronization break is observed when the chaotic attractors of both systems retain phase coherence on the plane $((\dot{x}, \dot{y})$ both below and above the phase synchronization boundary. For $a > a_*$, the chaotic attractor of the first system becomes phase-incoherent as soon as coupling parameter $\varepsilon$ turns out to be smaller than its value, $\varepsilon_{ps}$, at the phase synchronization boundary. As for the second system, its attractor remains phase-coherent. Note that $a_* > a_{c1,2}$; hence, the breakdown scenario changes when both coupled systems have phase-incoherent chaotic attractors on the plane $(x, y)$. The second type of phase synchronization break is shown in Fig. 5.

Importantly, the phase coherence of the chaotic attractor on the velocity plane $(\dot{x}, \dot{y})$ breaks much as this takes place on the coordinate plane $(x, y)$ of system (8) with initially phase-coherent attractors.

Thus, having studied the dynamics of two mutually coupled Ressler systems (described by (9)) with phase-incoherent attractors, we can distinguish two scenarios of synchronization breakdown, which are the same as those observed for oscillators with phase-incoherent attractors (see system (8)).

## 5. DYNAMICS OF TWO COUPLED TUNNEL-DIODE-BASED OSCILLATORS

Consider by way of example two mutually coupled tunnel-diode-based stochastic oscillators, which were described, e.g., in [31, 32]. The principal circuit of bidirectionally coupled oscillators is shown in Fig. 6. In the dimensionless form, the behavior of these self-oscillators is described by the equations

$$\dot{x}_{1,2} = \omega_{1,2}^2 [h(x_{1,2} - \varepsilon(y_{2,1} - y_{1,2})) + y_{1,2} - z_{1,2}],$$
$$\dot{y}_{1,2} = -x_{1,2} + \varepsilon(y_{2,1} - y_{1,2}), \qquad (10)$$
$$\mu \dot{z}_{1,2} = x_{1,2} - f(z_{1,2}),$$

where $x_{1,2} \sim I_{1,2}$, $y_{1,2} \sim U_{1,2}$, $z_{1,2} \sim V_{1,2}$, $h = MS/\sqrt{LC}$, $\mu = \tilde{C}/C$, $\omega_{1,2} = \sqrt{L/L_{1,2}}$, $\varepsilon = \sqrt{L}/(R\sqrt{C})$ is the coupling parameter, and $L$ is a normalizing factor. The control parameters were taken from [31]: $h = 0.2$ and $\mu = 0.1$. As dimensionless characteristic $f(\xi)$ of the nonlinear element, we used the function $f(\xi) = -\xi + 0.002sh(5\xi - 7.5) + 2.9$. The value of $\omega_2$ was fixed ($\omega_2 = 1.02$), while $\omega_1$ and $\varepsilon$ were varied.

Figure 7 shows the phase synchronization boundary and two characteristic curves $L_{1,2}$ specified by the dependences of the highest Lyapunov exponents of the system on coupling parameter $\varepsilon$. Since the oscillators are mutually coupled in our case, all six Lyapunov exponents depend on the coupling parameter. Yet, when the coupling parameter is zero, system (10), just as system (8), will be characterized by two zero Lyapunov exponents. As follows from Fig. 7, curve $L_1$ coincides with the phase synchronization boundary only in a very narrow range of the mismatch between the coupled oscillator parameters where control parameters $\omega_{1,2}$ are mismatched insignificantly. Thus, as for system (8), the passage of one of the zero Lyapunov exponents into the negative domain is a precursor to phase synchronization, rather than being its manifestation.

The position of curve $L_2$ on the plane ($\omega_1, \varepsilon$) is also totally unrelated to the phase synchronization boundary (Fig. 6). Consequently, the inferences regarding two unidirectionally coupled Ressler oscillators (system (8)) hold true for two bidirectionally (mutually) coupled oscillators (system (10)). In other words, two mutually coupled stochastic tunnel-diode-based self-oscillators exhibit the same two scenarios of phase synchronization breakdown as coupled Ressler oscillators: if the control parameter (frequency) mismatch is small, synchronization break is associated with the loss of the general rhythm of chaotic oscillations, while the attractors of both chaotic systems remain phase-coherent; if the frequency mismatch is large, the phase synchroni-zation breaks as one of the chaotic attractors loses phase coherence.

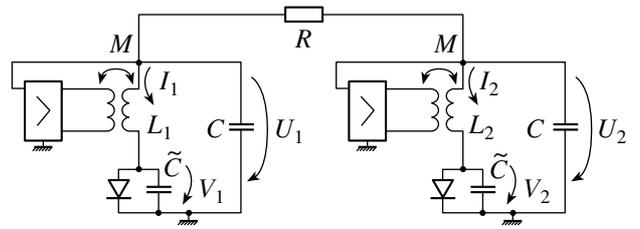

**Fig. 6.** Tunnel-diode-based oscillators with dissipative bidirectional coupling.

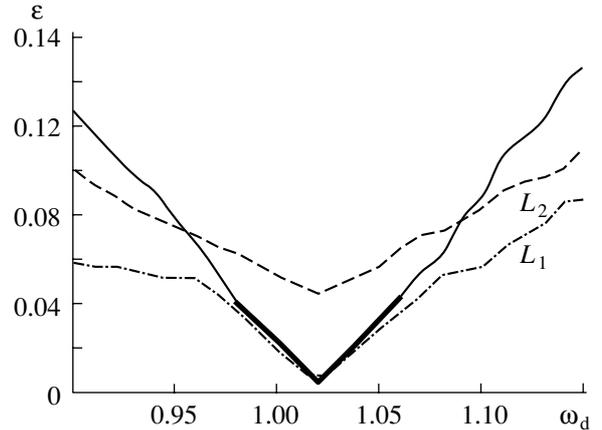

**Fig. 7.** Phase synchronization domain on the ($\omega_1, \varepsilon$) for system (10). For small mismatches $|\omega_1 - \omega_2|$, the phase synchronization boundary is shown by the thick line; for large mismatches, the boundary is shown by the thin line. Dash-and-dot line $L_1$ corresponds to the passage of one zero Lyapunov exponent into the negative domain, and dashed line $L_2$ shows vanishing of one positive Lyapunov exponent. The regions of the periodic and chaotic behavior are not discriminated.

## CONCLUSIONS

To summarize, we can state that chaotic oscillators feature two scenarios of phase synchronization breakdown. From the results obtained in this work, it follows that these two scenarios are similar to those observed for periodic oscillators in the case of classical synchronization. The first type of synchronization break is associated with the loss of the general rhythm of chaotic oscillations; the second, with the loss of phase coherence by chaotic attractors. These two scenarios are observed in systems with initially both phase-coherent and phase-incoherent chaotic attractors.

Another important outcome is that one of the zero Lyapunov exponents describing the behavior of coupled chaotic oscillators passes into the negative domain earlier than the phase synchronization regime sets in and, hence, in no way specifies the phase synchroniza-tion boundary. In just the same way, the phase synchronization boundary is totally unrelated to vanishing of one of the positive Lyapunov exponents. Therefore, description of phase synchronization in terms of the highest Lyapunov exponents seems to be incorrect.

ACKNOWLEDGMENTS

This work was supported by the Russian Foundation for Basic Research (project nos. 05-02-16273 and 06-02-16451), program in support of the leading scientific schools of the Russian Federation (grant no. NSh-4167.2006.2), Federal Agency of Science and Innovations (project nos. 2006-RI-19.0/001/053 and 2006-RI-19.0/001/054), CRDF (grant no. REC-006), and program "Dynasty."